\documentclass{llncs}
\usepackage{algorithm}
\usepackage{algorithmic}
\usepackage{graphicx}
\usepackage{url}
\usepackage{epstopdf}
\begin{document}
\frontmatter 
\pagestyle{headings} 
\author{Jijie Wang, Lei Lin, Ting Huang, Jingjing Wang and Zengyou He}
\institute{School of Software,\\Dalian University of Technology, \\Liaoning,China,116600,\\ Jijie.Wang@hotmail.com}
\mainmatter
\title{Efficient K-Nearest Neighbor Join Algorithms for High Dimensional Sparse Data}
\maketitle 
\begin{abstract}
The K-Nearest Neighbor (KNN) join is an expensive but important operation in many data mining algorithms. Several recent applications need to perform KNN join for high dimensional sparse data. Unfortunately, all existing KNN join algorithms are designed for low dimensional data. To fulfill this void, we investigate the KNN join problem for high dimensional sparse data.

\hspace{1em} In this paper, we propose three KNN join algorithms: a brute force (BF) algorithm, an inverted index-based(IIB) algorithm and an improved inverted index-based(IIIB) algorithm. Extensive experiments on both synthetic and real-world datasets were conducted to demonstrate the effectiveness of our algorithms  for high dimensional sparse data.
\end{abstract}
\section{Introduction}
The K-Nearest Neighbor (KNN) join operation associates each data point in one set $R$ with its $k$-nearest neighbors from another set $S$. Such KNN join operation can be used as a primitive building block in many data mining algorithms. Typical examples include $k$-nearest neighbor classification and $k$-means clustering. Therefore, several effective algorithms have been proposed recently \cite{ex2,ex3,ex4,ex6}. However, all these existing KNN join algorithms were designed for low dimensional data in which the dimensionality is generally less than 100. This deficiency restricts the use of the KNN join in practice since several recent applications in proteomics need to perform such operation for extremely high dimensional sparse data. 

Proteomics is a relatively new but rapidly developing concept within life science research \cite{ex11}. Peptide identification is the key and essential step in proteomics. In this process, the peptides in the sample are identified by searching a protein database according to the measured mass tandem mass spectra (MS/MS)\cite{ex12}.

From the view of computation, such MS/MS-based peptide identification problem can be boiled down to a KNN join problem: Set $R$ is the group of all experimental MS/MS spectra and set $S$ is a group of theoretic spectra derived from all peptides in the database, where each spectrum is a high dimensional sparse vector (the dimensionality is around 10,000). The objective is to find $k$ possible peptides that are associated with every input MS/MS spectrum. Unfortunately, all existing peptide identification algorithms have never optimized their search methods from a KNN join's perspective, resulting in considerable running deficiency.

Based on above observations, this paper focuses on the KNN join problem for high dimensional sparse data. To the best of our knowledge, there are still no available algorithms that are designed for the same purpose.

In this paper, we first propose a brute force algorithm as the baseline. Then an inverted index-based algorithm is proposed. This algorithm has significant performance benefits since it can avoid iterating unnecessary features in every vector of set $S$. In order to further improve the efficiency of the inverted index-based algorithm, we add a threshold-based refinement to it. This refinement can dramatically reduce overhead such as index construction and inverted list scanning during score accumulation.

Our contributions can be summarized as follows:
\begin{itemize}
\item We are first to investigate the high-dimensional KNN join problem for sparse vectors, which is rooted from the peptide identification problem in computational proteomics.
\item We propose several novel KNN join algorithms for sparse vectors in the high dimensional space.
\end{itemize}

The remainder of our paper is organized as follows: In section 2, we review some related work in KNN join algorithms. In section 3, we define the KNN join problem. Our proposed algorithms are presented in Section 4. Section 5 provides a performance evaluation based on both synthetic and real data. Finally, we conclude the paper in Section 6. 
\section{Related work}
Several KNN join algorithms have been proposed in the past a few years \cite{ex2,ex3,ex4,ex5,ex6}. MuX\cite{ex5,ex6} is essentially an R-tree based method designed to satisfy the conflicting requirements of reducing both CPU cost and I/O cost. It employs large-sized pages to optimize I/O time and uses small-sized buckets to partition the data with finer granularity so that CPU time can be reduced. Gorder\cite{ex4} is a block nested loop join method that exploits sorting, join scheduling and distance computation filtering to reduce both I/O and CPU costs. It sorts input datasets by G-order and applies the scheduled block nested loop join on the G-ordered data. Yu et al.\cite{ex2,ex3} propose an index structure called $iDistance$ to partition the data. Reference points are selected for each partition, and every point in each partition is mapped to a single dimensional space based on its similarity score to the corresponding reference point. Then the one-dimensional KNN search is performed on the transformed data indexed by a B+ tree. Since all these existing researches focus on data whose dimensionality is less than 100, none of them could be used to handle high dimensional sparse data with nearly 10000 features.

Bayardo et al.\cite{ex1} deal with high dimensional sparse data, but their objective is to find all pairs whose similarity score is above a threshold rather than finding each point with its $k$ nearest neighbors. Meanwhile, they focus on self similarity search and propose an efficient in-memory approach.

In our work, we devote ourselves to solving the KNN join problem with a large collection of sparse vector data in the high-dimensional space. Furthermore, the block nested loop join strategy is introduced to our algorithm since the size of the data set may exceed the available memory.
 
\section{Problem definition}
\begin{description}
\item[KNN join:] Given two data sets $R$ and $S$, an integer $k$ and the metric $sim()$, the KNN join of $R$ and $S$, denoted as ${R}\bowtie_{KNN}{S}$, returns pairs of vectors ($r$, $s$) ,where $r$ ${\in}$ $R$ and $s$ ${\in}$ $S$, and there are at most $k$-$1$ vectors from $S$ such that $sim(r,s')$${\ge}$$sim(r,s)$ (for any $s'$ ${\in}$  $S$).
\end{description}

In essence, the KNN join combines each vector in outer dataset $R$ with its $k$-nearest neighbors in inner dataset $S$. Both $R = \{r_{1},r_{2},...,r_{|R|}\}$ and $S = \{s_{1},s_{2},...,s_{|S|}\}$ are composed of real-valued vectors of fixed dimensionality $D$, where $|R|$ and $|S|$ represent the number of vectors in set $R$ and $S$ respectively. 

In this paper, the similarity metric is the ${dot\ product}$. For each vector $r$ in $R$ and $s$ in $S$, we calculate their similarity score as follows:
\begin{equation}
dot(r,s) = \sum_{i=1}^{D} r[i]{\cdot}s[i]
\end{equation}

Due to the sparsity of input vectors, we present a sparse vector $x$ as a set of pairs ($d$,$w$), where $d$ is the dimension index iterator and $w = x[d]$, $w > 0$ over all $d$ = 1...$D$. In general, all these pairs are organized in ascending order of the value $d$. Like \cite{ex1}, we define such pairs as the $features$ of the vector and the size of these $features$ as $|x|$.
\section{Algorithms}
In this section, we describe our proposed algorithms. We first describe the block nested loop join in our work. Then a brute force algorithm is introduced as a baseline in Section 4.2. In order to improve the efficiency of brute force algorithm, we propose an inverted index-based algorithm and an improved inverted index-based algorithm in Section 4.3 and 4.4, respectively. 
\subsection{Block nested loop join}
Block nested loop join is designed for effectively utilizing buffer pages to deal with disk resident data. It uses one page as the input buffer for scanning the inner data set $S$, one page as the output buffer, and uses all remaining pages to hold blocks of outer $R$. It has been proved that this strategy of caching more pages of the outer set is beneficial for I/O processing, because the inner set can be scanned for fewer times\cite{ex9}.
In our KNN join problem, the sizes of data set $R$ and $S$ tend to exceed the available memory. Therefore we introduce block nested loop strategy into our algorithm. Suppose we allocate $n_{r}$ and $n_{s}$ buffer pages for $R$ and $S$. We partition $R$ and $S$ into blocks with the allocated buffer size. Both the blocks of $R$ and $S$ are loaded into memory sequentially, which is efficient in terms of I/O time as it significantly reduces seek overhead.

Algorithm 1 outlines the block nested loop join algorithm. It loads blocks of data set $R$ into memory sequentially (line 1-2). For the in-memory block $B_{r}$, $pruneScore$ of its vectors is initialized to 0 (line 3). Then blocks of data set $S$ are loaded into memory one by one (line 4-5). With each pair $B_{r}$ and $B_{s}$, we join them in memory by calling function $KNN\_Join\_Algorithm$ (line 6). After computing the similarity score between vectors from $B_{r}$ and every block in data set $S$, the KNN candidate sets for every vector in $B_{r}$ will be outputted as the join results (line 7).

\begin{algorithm}[ht!]
\caption{Block\_Nested\_Loops\_Join($R$, $S$)}
\begin{description}
\item[Input:] $R$, $S$ are two sparse vector data sets that have been partitioned into blocks.
\item[Description:]
\end{description}
\begin{algorithmic}[1]
\STATE For each block $B_{r}$ ${\in}$ $R$ do 
\STATE ${\quad}$ ReadBlock(${B_{r}}$)
\STATE ${\quad}$ InitPruneScore(${B_{r}}$)
\STATE ${\quad}$ For each block $B_{s}$ ${\in}$ $S$ do
\STATE ${\qquad}$ ${\quad}$ReadBlock(${B_{s}}$)
\STATE ${\qquad}$ ${\quad}$KNN\_Join\_Algorithm (${B_{r}}$,${B_{s}}$)
\STATE ${\quad}$ OutputKNN(${B_{r}}$)
\end{algorithmic}
\end{algorithm}
\subsection{Brute force algorithm}
As a baseline, one might consider a brute force algorithm: simply compute the similarity score between every vector $r$ in ${B_{r}}$ and every vector $s$ in ${B_{s}}$. If the similarity score is higher than the pruning score of $r$, then $s$ is inserted into $r$'s KNN candidate set and at the same time the pruning score is updated.

Algorithm 2 shows the outline of our brute force (BF) algorithm for KNN join problem. In BF algorithm, the most important part is the function $dot(r,s)$. This function could be implemented using a fast algorithm (shown in line 8-23) and its complexity is:
\begin{equation}
{C_{1}} = |r|+|s|,
\end{equation}
where $|r|$ refers to the number of features in vector $r$ and $|s|$ refers to the number of features in vector $s$. Then, we calculate the complexity of BF algorithm as follows:
\begin{equation}
{C_{2}} = \sum_{i=1}^{ |B_{r}|} \sum_{j=1}^{|B_{s}|} \{|r_{i}|+|s_{j}| \},
\end{equation}
where ${|B_{r}|}$ and ${|B_{s}|}$ are the number of vectors in ${B_{r}}$ and ${B_{s}}$ respectively.
\begin{algorithm}[ht!]
\caption{Brute Force (BF) Algorithm}
\begin{algorithmic}[1]
\STATE KNN\_Join\_Algorithm\_BF(${B_{r}}$,${B_{s}}$): 
\STATE For each vector $r$ ${\in}$ ${B_{r}}$ do 
\STATE ${\quad}$ For each vector $s$ ${\in}$ ${B_{s}}$ do
\STATE ${\quad}$ v = dot($r$, $s$)
\STATE ${\quad}$ If (v$>$ pruneScore($r$)) then 
\STATE ${\qquad}$ ${\quad}$ Insert $s$ into the KNN candidate set of $r$
\STATE ${\qquad}$ ${\quad}$ Update pruneScore($r$)
\\ $*$pruneScore($r$) represents the similarity score between vector $r$ and its $k$th nearest neighbor 
\\[10pt]
\STATE dot($r$, ${s}$):
\STATE ret=0
\STATE iterator\_r = iterator that traverses through all the features in $r$
\STATE iterator\_s = iterator that traverses through all the features in $s$
\STATE While (iterator\_r.hasNext() and iterator\_s.hasNext()) do 
\STATE ${\quad}$ feature\_r = iterator\_r.currentValue() 
\STATE ${\quad}$ feature\_s = iterator\_s.currentValue()
\STATE ${\quad}$ If (feature\_r.$d$$==$feature\_s.$d$) then
\STATE ${\qquad}$ ${\quad}$ ret = ret + feature\_r.$w$${\cdot}$feature\_s.$w$
\STATE ${\qquad}$ ${\quad}$ iterator\_r.next()
\STATE ${\qquad}$ ${\quad}$ iterator\_s.next()
\STATE ${\quad}$ Else If (feature\_r.$d$ $>$ feature\_s.$d$) then
\STATE ${\qquad}$ ${\quad}$ iterator\_s.next()
\STATE ${\quad}$ Else 
\STATE ${\qquad}$ ${\quad}$ iterator\_r.next()
\STATE Return ret
\end{algorithmic}
\end{algorithm}
\subsection{Inverted index-based algorithm}
In our work, we deal with a large collection of sparse vectors in the high dimensional space. Hence, the efficiency of BF algorithm is far from satisfactory. One reason for BF's low efficiency is due to its traversal over a great number of unnecessary features in vector $s$ during the calculation of $dot(r,s)$. However, these features have no contributions to the similarity score of $r$ and $s$. This problem will still exist even we use other methods, such as hashing and indexing, to facilitate the implementation of $dot(r,s)$. Considering this fact, we introduce $inverted$ $list$ into our algorithm.

$Inverted$ $list$ is a set of lists \{${I_{1},I_{2},...,I_{D}}$\} (one for each dimension). Each list $I_{d}$ consists of a set of pairs $(x,w)$, where $x$ ${\in}$ $S$, $w$$=$$x[d]$ and $w$ is non-zero. 

Algorithm 3 shows our inverted index-based algorithm: IIB. First, it calls function Create\_Inverted\_List\_IIB to create inverted list (line 2). Then it traverses every vector $r$ in ${B_{r}}$ and calls function Find\_Matches\_IIB to find $k$-nearest neighbors of $r$ (line 3-4).

Line 5-8 outlines Create\_Inverted\_List\_IIB function. This function first visits every vector $s$ in ${B_{s}}$ (line 6). And for every feature in ${s}$, it inserts the pair ($s,s[d]$) into $I_{d}$(line 7-8).

Find\_Matches\_IIB is shown in line 9-17. First, it constructs a map $A$ to connect vector id and similarity value (line 10). And then for every feature($d,r[d]$) in $r$, it traverses corresponding $I_{d}$ and add $r[d]$ ${\cdot}$ $s[d]$ to $A[s]$ for every existing pair($s,s[d]$) in $I_{d}$ (line 11-13). After all elements in $A$ have been visited, the $k$-nearest neighbors of vector $r$ are found. Note that when the KNN candidate set of $r$ is updated, pruneScore($r$) should also be updated.

Now, we compute the complexity of IIB algorithm:
\begin{equation}
{C_{3}} = \sum_{i=1}^{ |B_{s}|} |s_{i}| + \sum_{i=1}^{ |B_{r}|} \sum_{j=1}^{|r|} {|I_{r[j].d}| },
\end{equation}
where $r$ refers to the $i$th vector in $B_{r}$ and $|I_{r[j].d}|$ refers to the number of pairs ($s,w$) in $I_{r[j].d}$.

Compared with $C_{2}$, $C_{3}$ dramatically drops because IIB algorithm avoids iterating many unnecessary features in $s$ for the corresponding vector $r$. 
\begin{algorithm}[ht!]
\caption{Inverted Index-based (IIB) Algorithm}
\begin{algorithmic}[1]
\STATE KNN\_Join\_Algorithm\_IIB(${B_{r}}$,${B_{s}}$):
\STATE Create\_Inverted\_List\_IIB(${B_{s}}$)
\STATE For each vector $r$ ${\in}$ ${B_{r}}$ do 
\STATE ${\quad}$ Find\_Matches\_IIB(${{r}}$)
\\[10pt]
\STATE Create\_Inverted\_List\_IIB(${B_{s}}$):
\STATE For each vector $s$ ${\in}$ ${B_{s}}$ do 
\STATE ${\quad}$ For each feature ($d$, $w$) ${\in}$ $s$ do
\STATE ${\qquad}$ ${\quad}$ Insert pair ($s$, $w$) into list $I_{d}$
\\[10pt]
\STATE Find\_Matches\_IIB(${{r}}$):
\STATE $A$ = empty map from vector to similarity score
\STATE For each feature ($d$, $r[d]$) ${\in}$ $r$ do
\STATE ${\quad}$ For each pair ($s$, $s[d]$) ${\in}$ $I_{d}$ do
\STATE ${\qquad}$ ${\quad}$ $A[s]$ = $A[s]$ + $r[d]$ ${\cdot}$ $s[d]$ 
\STATE For each $s$ with non-zero score ${\in}$ $A$ do
\STATE ${\quad}$ If ($A[s]$ $>$ pruneScore($r$)) then
\STATE ${\qquad}$ ${\quad}$ Insert $s$ into the KNN candidate set of $r$ 
\STATE ${\qquad}$ ${\quad}$ Update pruneScore($r$)
\end{algorithmic}
\end{algorithm}
\subsection{Improved inverted index-based algorithm}
Many researches \cite{ex1,ex8,ex10} on distance join utilize the similarity score threshold to determine which candidate pair should be added into the result set. Roberto et al.\cite{ex1} go a step further to exploit such threshold to reduce the amount of information indexed in the inverted lists. Unlike distance join, KNN join does not have a pre-determined threshold and hence we cannot directly utilize existing algorithms on similarity join problem. Thanks to the specific property of block nested loop join, we could utilize the computation results from previous loops to obtain a threshold that could be used in forthcoming loops. That is, we define the minimum similarity score in current block ${B_{r}}$ as $MinPruneScore$ = $min_{r\in B_{r}}{pruneScore(r)}$. Note that when we finish joining block ${B_{r}}$ and previous blocks from $S$, $pruneSocre(r)$ is updated and so does $MinPruneScore$. Hence, $MinPruneScore$, as a threshold derived from previous computation, could be used to help compute the similarity scores between block ${B_{r}}$ and current block ${B_{s}}$ in $S$. 

This threshold-based refinement is added to our improved inverted index-based algorithm, as shown in Algorithm 4. The loop in Create\_Inverted\_List\_IIIB now iterates from the most frequent feature to the least one (line 6-10), and avoids inserting any pair($s,s[d]$) into the $I_{d}$ until a condition is met(line 11-13). The rationale is that we only need to include potential KNN candidates of $r$ into the inverted lists. The frequency-based ordering aims at minimizing the length of the inverted lists, which has also been utilized in {\cite{ex1}}.

In order to preserve memory and speed up the computation, we remove the feature($d,w$) from $s$ after it is inserted into $I_{d}$ (line 14). Because only partial features are indexed in the inverted lists, Find\_Matches\_IIIB(line 17-19) does not finish the score accumulation task.  Therefore, we need to continue the score calculation using non-indexed features(line 21). After doing these, we can get an exact similarity score between two vectors. 

\begin{theorem}
If there exist two vectors $r$ and $s$ such that dot($r,s$) $>$ pruneScore($r$), then $s$ will become one of the KNN candidates of $r$.
\end{theorem}
\begin{proof}
We use $s'$ to denote the vector with unindexed features of $s$ and $s''$ to denote the vector with indexed features of $s$. In Create\_Inverted\_List\_IIIB, when a trivial upper bound $t$ exceeds MinPruneScore, it will begin to index the remaining features. Thus for any vector $r$ and indexed vector $s''$, there exists dot($r,s'$) $<$ MinPruneScore. Since pruneScore($r$) ${\ge}$ MinPruneScore and dot($r,s$) $>$ pruneScore($r$), we can deduce that dot($r,s$) $>$ MinPruneScore. Note that dot($r,s$) = dot($r,s'$) + dot($r,s''$), hence for any vector $r$ and unindexed vector $s'$ meeting the condition dot($r,s'$) $<$ MinPruneScore, we have dot($r,s''$) $>$ 0. Therefore, we have at least one indexed feature of $s$ in common with one feature of $r$, and the similarity score of $r$ and $s$ will be computed completely in line 21. After computing the similarity score and getting dot($r,s$) $>$ pruneScore($r$), $s$ will become one of the KNN candidates of $r$. 
\qed
\end{proof}
\begin{algorithm}[ht!]
\caption{Improved Inverted Index-based (IIIB) Algorithm}
\begin{algorithmic}[1]
\STATE KNN\_Join\_Algorithm\_IIIB(${B_{r}}$,${B_{s}}$):
\STATE Create\_Inverted\_List\_IIIB(${B_{s}}$)
\STATE For each vector $r$ ${\in}$ ${B_{r}}$ do 
\STATE ${\quad}$ Find\_Matches\_IIIB($r$)
\\[10pt]
\STATE Create\_Inverted\_List\_IIIB(${B_{s}}$):
\STATE Reorder the dimension 1...$D$ such that dimensions with the more non-zero entries in $B_{r}$ appear first
\STATE Denote the maximum value $x[d]$ over all $x$ ${\in}$ ${B_{r}}$ as $maxWeight_{d}({B_{r}})$.
\STATE For each vector $s$ ${\in}$ ${B_{s}}$ do 
\STATE ${\quad}$ $t$ = 0;
\STATE ${\quad}$ For each feature ($d$, $w$) ${\in}$ $s$ do
\STATE ${\qquad}$ ${\quad}$ $t$ = $t$ + $maxWeight_{d}$($B_{r}$) ${\cdot}$ $w$
\STATE ${\qquad}$ ${\quad}$ If ( t $>$ MinPruneScore ) then
\STATE ${\qquad}$ ${\quad}$${\qquad}$ Insert pair ($s$, $w$) into $I_{d}$
\STATE ${\qquad}$ ${\quad}$${\qquad}$ Remove feature ($d$, $w$) from $s$
\\[10pt]
\STATE Find\_Matches\_IIIB(${r}$):
\STATE $A$ = empty map from vector to similarity score
\STATE For each feature ($d$, $r[d]$) ${\in}$ $r$ do
\STATE ${\quad}$ For each pair ($s$, $s[d]$) ${\in}$ $I_{d}$ do
\STATE ${\qquad}$ ${\quad}$ $A[s]$ = $A[s]$ + $r[d]{\cdot}s[d]$ 
\STATE For each s with non-zero score ${\in}$ $A$ do
\STATE ${\quad}$ $A[s]$ = $A[s]$ + dot($r, s$) 
\STATE ${\quad}$ If ($A[s]$ $>$ pruneScore($r$)) then
\STATE ${\qquad}$ ${\quad}$ Insert $s$ into the KNN candidate set of $r$ 
\STATE ${\qquad}$ ${\quad}$ Update pruneScore($r$)
\end{algorithmic}
\end{algorithm}
\section{Experimental Results}
A number of experiments were conducted to evaluate the performance of our KNN join algorithms. The data sets used in the evaluation consist of both synthetic and real-world data sets. Our synthetic data sets consist of 10,000 to 100,000 random sparse vectors with 10,000 dimensions. Our real data is the MS/MS data obtained from spectra of the Yeast and Worm (\url{http://noble.gs.washington.edu/proj/percolator/}). 

In the Yeast and Worm datasets, each spectrum acts as the sparse vector in our experiments, where Yeast consists of 35,236 vectors and Worm consists of 207,804 vectors. In each spectrum, we treat each peak as the feature of vector. In the pre-processing step, the value of $m/z$ multiplies 10 to serve as dimension index and peak's intensity is directly used as the value of corresponding dimension. In our experiments, we use the Yeast as dataset $R$ and the Worm as dataset $S$. 

The experiments were performed on a 2.4 GHz machine with 2G RAM and a 7200 RPM SATA-IDE hard disk. The default settings of our experiments are summarized in Table 1.

We implement all three KNN join algorithms: BF, IIB and IIIB and compare their performance from different perspectives. Since those three algorithms utilize the same block nested loop strategy to handle the disk resident data, we only choose the I/O time of BF to show our algorithms' I/O performance. Furthermore, the performances of three algorithms in terms of the CPU time are presented.

\begin{table}[h]
\caption{Default parameter values} \label{table1}
\begin{center}
\begin{tabular}{|c|c|c|c|}
\hline
\ Parameter & Default Setting \\
\hline
\ Number of nearest neighbors & 5 \\
\hline
\ Buffer size& Around 50\% of total size of $R$ and $S$ \\
\hline
\ Size of $R$ data in buffer & Around 80\% of buffer \\
\hline
\ Buffer page size & 8192 \\
\hline
\end{tabular}
\end{center}
\end{table}
\subsection{Evaluation using synthetic datasets}
In this set of experiments, we compare the performance of BF, IIB and IIIB algorithms using the synthetic dataset.
\subsubsection{Effect of data size}
We first study the effect of varying data size on these three algorithms. Fig.1 shows the results for KNN join on 10000-dimensional synthetic datasets of size varying from 10,000 to 50,000. From the results, we observe that with the increase of data size, the cost of BF increases dramatically, while IIIB and IIB perform more stable and better than BF. That is because IIB could avoid iterating unnecessary features in data set $S$, and the threshold-based refinement in IIB could prune more vectors. 
\begin{figure}[h]
\centerline{\includegraphics[scale=0.4]{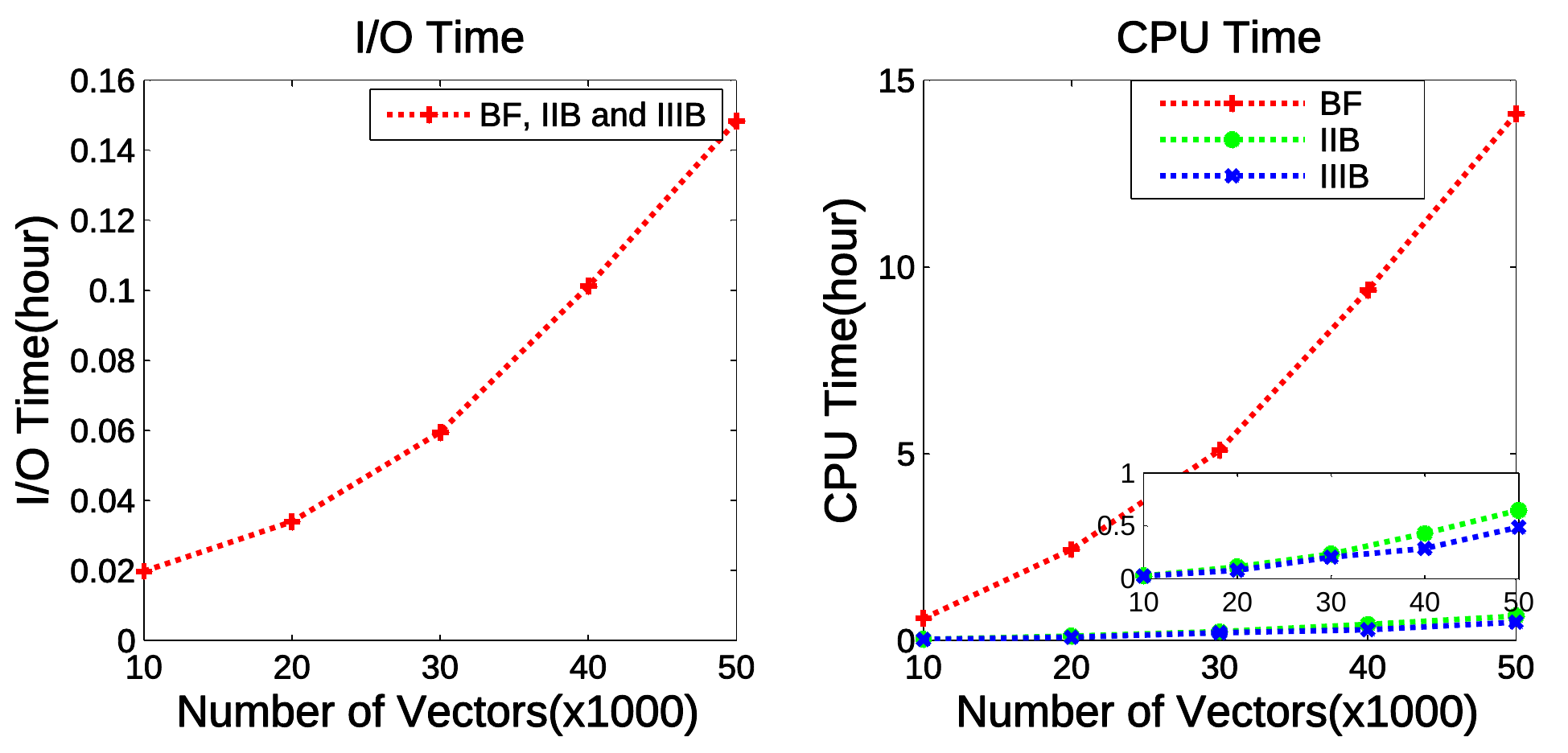}}
\caption{Effect of data size (10000-dimensional synthetic datasets ) }
\end{figure}
\subsubsection{Effect of relative size of data sets}
In this set of experiments, we joined two datasets of different sizes and studied the effect of the relative size on the performance of the join algorithm. To study such an effect, we fixed the size of $R$ at 10,000 vectors and varied the size of $S$ from 1,000 to 100,000 so that the relative size of $R$:$S$ is changed from 10:1 to 1:10. Fig.2 shows the results. 

From the results, we can observe that the costs of BF, IIB and IIIB increase in proportion to the increase of the size of data set and are not heavily affected by the relative size. Meanwhile, IIIB is the most efficient algorithm compared with BF and IIB. 
\begin{figure}[h]
\centerline{\includegraphics[scale=0.4]{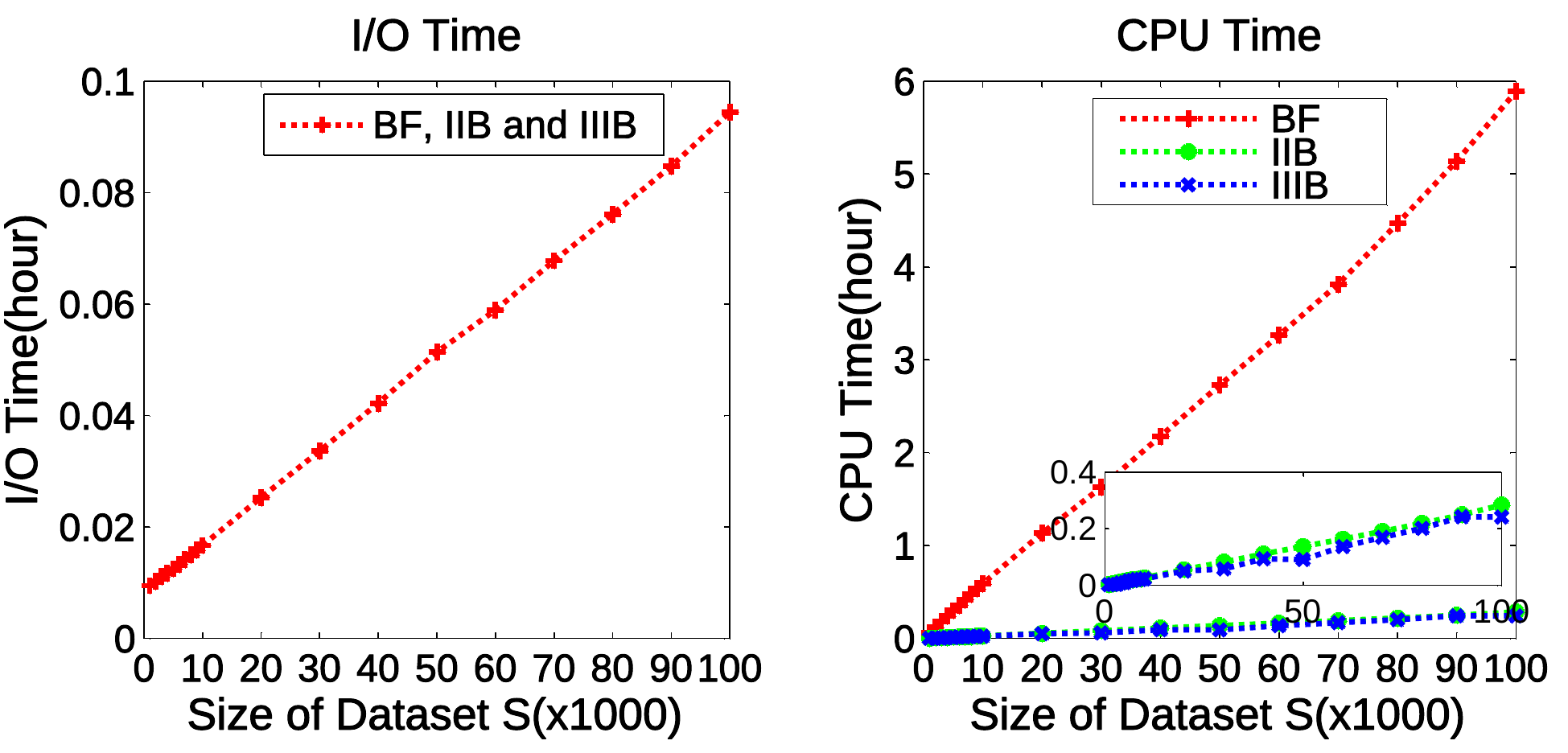}}
\caption{Effect of relative size of data sets (10000-dimensional synthetic datasets ) }
\end{figure}
\subsection{Evaluation using real datasets}
In this set of experiments, we study the performance of BF, IIB and IIIB algorithms using the real-world dataset.
\subsubsection{Effect of k} 
In this experiment, we study the effect of $k$. Fig.3 shows the performance results of three KNN join algorithms when $k$ is varied from 5 to 20 on the Yeast\&Worm datasets.

From the results, we notice that with the increase of the number of nearest neighbors, the I/O time almost remains unchanged. From the CPU time's perspective, all these three algorithms increase moderately because their pruning strategies do not rely on $k$. Thus the increase of $k$ just leads to generating more candidates for each vector, which will cost a bit more running time. On average, IIIB is about 16\% better than IIB for the Yeast\&Worm datasets, and both IIIB and IIB outperform BF with the speed-up factor of around 10. 
\begin{figure}[h]
\centerline{\includegraphics[scale=0.4]{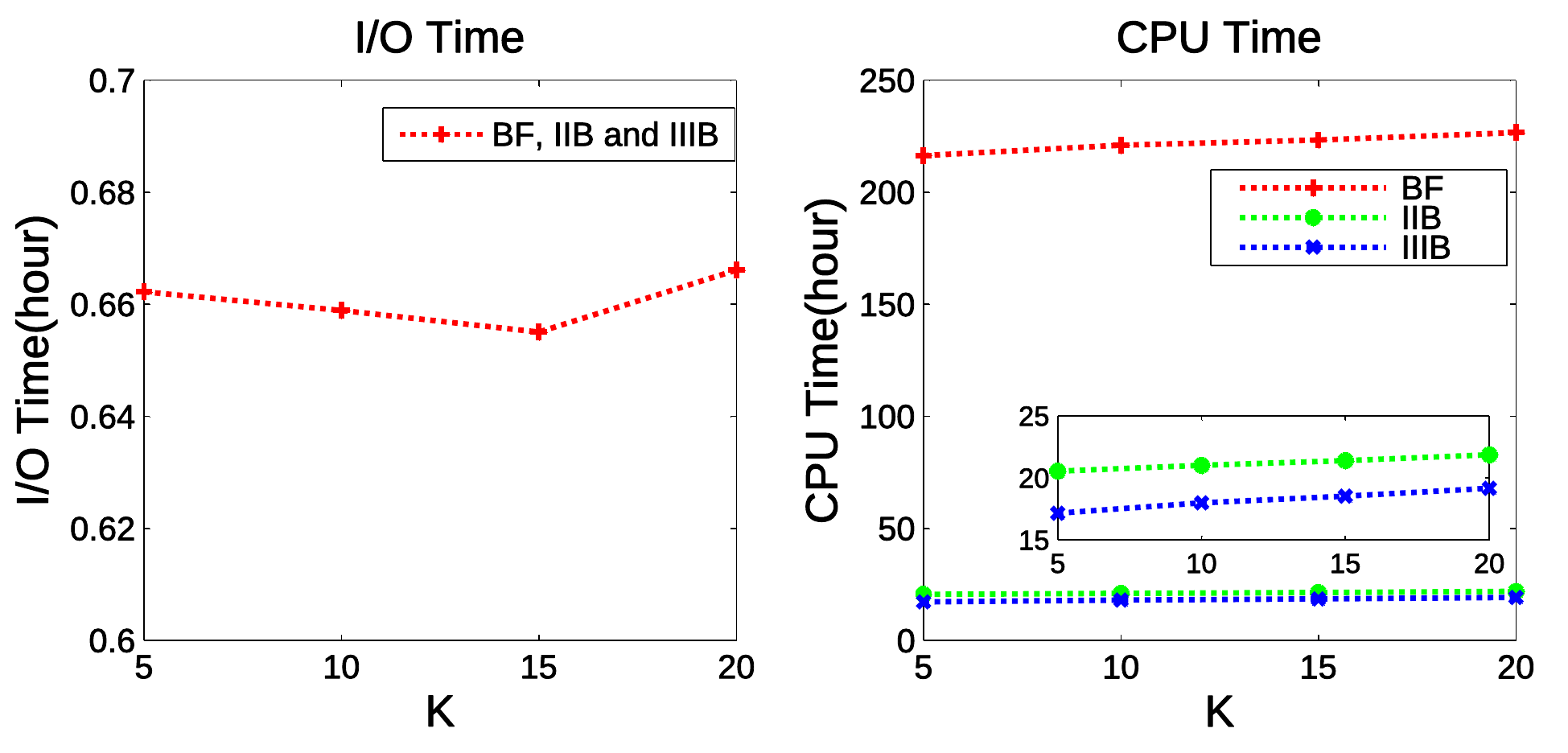}}
\caption{Effect of $k$ on Yeast\&Worm datasets}
\end{figure}
\subsubsection{Effect of buffer size}
In dealing with large datasets, the KNN join algorithm must be efficient in utilizing the limited buffer size. In this experiment, we study the behavior of the join methods with respect to buffer size.

The study is performed on the Yeast\&Worm datasets and we decrease the buffer size from around 50\% of the total dataset size to around 10\% of the total dataset size. In Fig.4, we compare the performance of BF, IIB and IIIB. From I/O time's perspective, with the decrease of the buffer size, the I/O time increases unavoidably because small buffer size will cost more I/O access. From CPU time's perspective, IIIB still performs better than IIB and the gap is more evident with the decrease of the buffer size. The reason is that smaller buffer size will bring a more powerful and accurate threshold-based refinement. Meanwhile, IIIB and IIB still have a great advantage over BF. 
\begin{figure}[h]
\centerline{\includegraphics[scale=0.4]{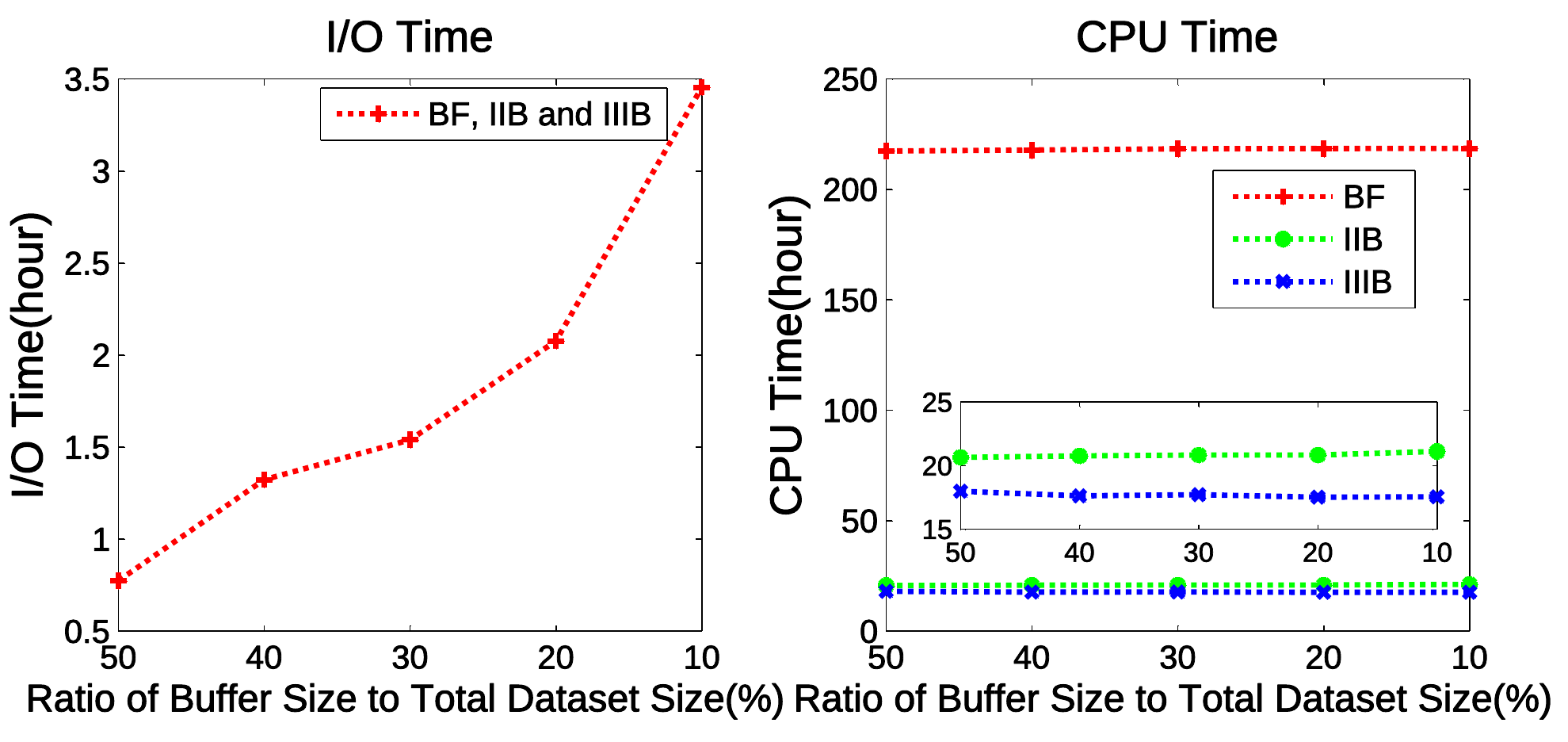}}
\caption{Effect of buffer size on Yeast\&Worm datasets }
\end{figure}
\section{Conclusions}
K-nearest neighbor join is the basis of many applications, including some recent applications in proteomics. These applications need to perform KNN join for extremely high dimensional sparse data. In this paper, we have proposed three algorithms to efficiently solve this problem. BF is a brute force algorithm which acts as the baseline. IIB utilizes the inverted lists and has a great advantage against BF. IIIB, which is based on IIB, uses the threshold-based refinement to solve the KNN join problem more efficiently. We did our performance study on both synthetic and real-world datasets. The results confirm that IIIB and IIB are scalable with respect to the number of nearest neighbors, buffer size, data size and relative size of dataset.

In the future work, we will focus on how to improve the efficiency of IIIB algorithm by finding more powerful refinement strategy. Meanwhile, we will use the proposed algorithms as the basis to implement a more efficient protein search engine.
\bibliographystyle{splncs}
\bibliography{IJoin}
\end{document}